\documentstyle[12pt,epsfig]{article}

 \newlength{\dinwidth}
 \newlength{\dinmargin}
\begin{document}

\newcommand{\be}{\begin{equation}}
\newcommand{\ee}{\end{equation}}
\newcommand{\tdm}[1]{\mbox{\boldmath $#1$}}  
\newcommand{\ie}{{\it i.e.\ }}
\newcommand{\ppbar}{$p\bar{p}$}
\newcommand{\pythia}{{\sc Pythia}}
\def\lesssim{\mathrel{\rlap{\lower4pt\hbox{\hskip1pt$\sim$}}
    \raise1pt\hbox{$<$}}}         
\def\gtrsim{\mathrel{\rlap{\lower4pt\hbox{\hskip1pt$\sim$}}
    \raise1pt\hbox{$>$}}}         

\titlepage
\begin{flushright}
TSL/ISV-2001-0255
\end{flushright}\begin{center}
\vspace*{2cm}
{\Large \bf 
Hard colour singlet exchange and \\
gaps between jets at the Tevatron
}  

\vspace*{1cm}
R.\ Enberg$^{a}$, G.\ Ingelman$^{a,b}$ and  L.\ Motyka$^{a,c}$ \\
\vspace*{0.5cm}
$^{a}$ High Energy Physics, Uppsala University, 
       Box 535, S-75121 Uppsala, Sweden   \\
$^{b}$ Deutsches Elektronen-Synchrotron DESY, Notkestrasse 85, 
       D-22603 Hamburg, Germany \\
$^{c}$ Institute of Physics, Jagellonian University, Reymonta 4, 
       30-059 Krak\'{o}w, Poland \\

\end{center}

\vspace*{1cm}

\begin{abstract}
The new kind of events with a rapidity gap between two high-$E_T$ jets,
observed in high energy $p\bar{p}$ collisions at the Tevatron, is found to be
well described by the exchange of a colour singlet gluon system in the BFKL
framework. This requires going beyond the conventional asymptotic 
Mueller-Tang approximation, which results in qualitatively different features of
 the basic parton-parton scattering amplitude. Non-leading corrections to
the BFKL equation are included by incorporation of the consistency constraint
and the running QCD coupling. Hadronisation and other non-perturbative QCD
effects are treated through a complete Monte Carlo simulation, providing a gap
survival probability that varies event-by-event, facilitating comparison
with experimental results. 
\end{abstract}

\newpage

\section{Introduction}

The phenomenon of rapidity gaps, \ie a region in rapidity or polar angle 
with no final state particles, has received much interest in recent years. 
This goes back to conventional diffractive scattering where a beam proton is 
quasi-elastically scattered, keeping most of its original momentum 
($x_F\gtrsim 0.9$), and
emerges separated by a rapidity gap from the remaining final state particles. 
The introduction of a hard scale in the process \cite{IS} has opened new 
possibilities to investigate an underlying parton process using perturbative 
QCD (pQCD). In spite of the hard scale, the rapidity gap is here always 
produced through soft, non-perturbative QCD dynamics as described in the 
pomeron model \cite{IS} or the soft colour interaction model \cite{SCI}. 
This is given by the fact that the forward proton  (or small mass system) 
is not affected by any large momentum transfer.  

\begin{figure}[hbtp]
\begin{center}
\begin{tabular}{cc} 
\begin{tabular}{c} 
\epsfig{width=0.3 \columnwidth, file=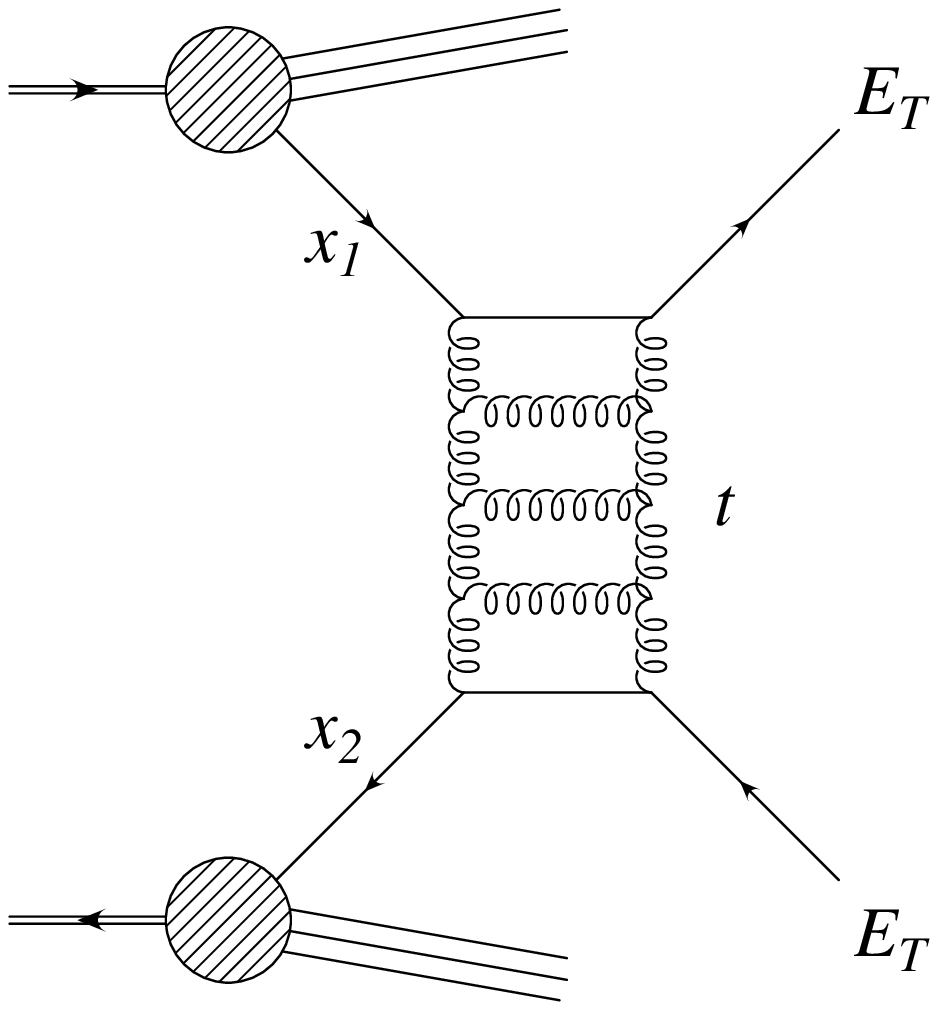} \\[14mm]
a)
\end{tabular}
&
\begin{tabular}{c} 
\hspace*{10mm}
\epsfig{width=0.17 \columnwidth,file=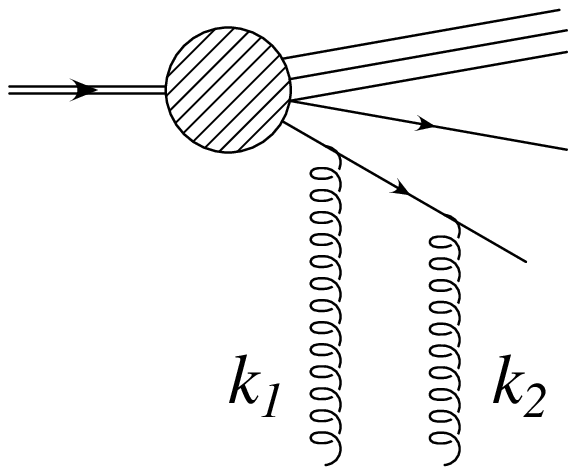} \\[4mm]
b) \\[10mm]
\hspace*{10mm}
\epsfig{width=0.17 \columnwidth,file=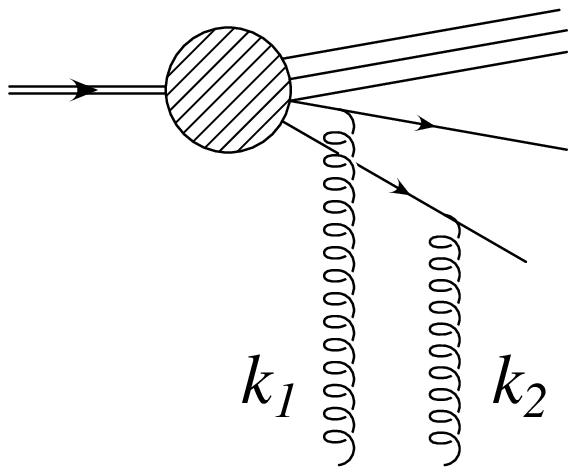} \\[4mm]
c) \\
\end{tabular}
\end{tabular}
\end{center}
\caption{(a) The jet-gap-jet event topology in $p\bar p$ scattering resulting
from hard colour singlet gluon ladder exchange between two partons as described
by the BFKL equation. 
(b,c) The impact factors for the pomeron coupling to the same parton and to two different partons.}
\label{fig1}
\end{figure}

A new class of rapidity gap events has recently been discovered 
in \ppbar \ collisions at
the Fermilab Tevatron \cite{D0DATA,CDFDATA}. The gap is then central 
in rapidity and has a high-$E_T$ jet on each side
corresponding to a large momentum transfer
$t\simeq - E^2_{T jet}$ across the gap. 
The models for soft gap formation do not give a satisfactory description of 
this new phenomenon \cite{SCI-jgj}, which calls for an explanation in terms of 
a hard colour singlet (HCS) exchange in terms of pQCD. 
Mueller and Tang \cite{MT} proposed the use
of the BFKL equation \cite{BFKL} to describe the exchange of a colour singlet 
gluon ladder, as illustrated in Fig.\ \ref{fig1}a, giving a rapidity gap between
the two hard-scattered partons. Their asymptotic solution does not, however,
give a good description of the experimental data \cite{D0DATA,CDFDATA}.

In this paper we provide a more complete, non-asymptotic solution to the BFKL equation 
for this process including also formally sub-leading effects. 
By implementing it in the {\sc Pythia} Monte Carlo \cite{Pythia} we take into account 
the soft QCD processes that are necessary to produce the hadronic final state. 
The soft processes can destroy the parton level gap and result in a gap 
survival probability significantly smaller than unity.  
Our complete model gives a good description of the observed `jet-gap-jet' events. 
This provides an important test of the BFKL equation which describes novel QCD dynamics,
beyond the conventional DGLAP parton evolution \cite{DGLAP}. 

\section{The BFKL calculation}

The leading logarithmic (LL) BFKL equation \cite{BFKL} was introduced 
to describe the total cross-section for scattering of two small objects 
by summing, in the leading $\log (1/x)$ approximation, gluonic ladder diagrams 
with reggeised gluons along the ladder. 
The BFKL equation naturally generates the steep rise at small $x$ of the gluon density in 
the proton and is of importance for the present exploration of QCD dynamics at HERA as well as 
for hadron colliders. The LL BFKL equation gives, however, a too steep energy dependence 
of the scattering amplitude and suffers from consistency problems due to the diffusion of
gluon momenta into the infrared domain. 
Substantial progress in how to cure these problems has been achieved in the last few years. 
The next-to-leading logarithmic (NLL) corrections to the BFKL kernel have been computed 
\cite{BFKLNL} and, after a proper treatment of the collinear divergencies \cite{COLLINEAR},
result in a lower intercept of the hard pomeron (about 1.2--1.3) which 
agrees with the experimental value. 
The infrared problems seem to be greatly reduced by unitarisation corrections to the
BFKL pomeron \cite{BalKov,GBMS}, which suppress the evolution for low momenta by adding a 
non-linear term in the BFKL equation.
The resulting phenomenon of saturation is confirmed by the success of the saturation model 
\cite{SATURATION}. 
Thus, it is important to include these theoretical improvements when making critical 
tests of the BFKL equation against experimental data.  

The BFKL formalism is also capable of taking into account non-forward hard colour 
singlet exchange, which plays a crucial role in high-$t$ elastic vector meson production
\cite{VMBFKL} and in events with gaps between jets \cite{MT,CFL}. 
The latter phenomenon is here studied in a theoretically improved BFKL treatment of the 
complex gluonic object constituting the hard pomeron. 
For asymptotically large rapidity separation $y$ between the jets, 
it has been shown \cite{MT,Bartels} that the LL~BFKL pomeron may be regarded as 
coupling to individual partons resolved at the scale $|t|$. 
The origin of this factorisation property which should
hold beyond the LL approximation, is motivated as follows.

When the rapidity $y$ is small the BFKL ladder (fig.\ \ref{fig1}a)
reduces to two simple gluons with transverse momenta $\tdm k_{1,2}$, with
$(\tdm k_{1} + \tdm k_{2})^2 \simeq -t$.
In the two gluon approximation there appear contributions from scatterings 
on one parton
(fig.\ \ref{fig1}b) and two partons (fig.\ \ref{fig1}c) in each proton. 
The contributions of individual diagrams involving the subdiagrams 
(figs.\ \ref{fig1}bc) are infrared divergent when $\tdm k_{1,2} \to 0$ . 
However, in the sum of all the contributions these divergencies cancel 
due to colour coherence and the colour neutrality of the proton and the final state. 
When both the exchanged gluons couple to the same pair of partons (as in fig.\ \ref{fig1}a), 
the loop integration gives a large logarithm  of the momentum transfer
while in the other case  (when subdiagram fig.\ \ref{fig1}c is relevant)
the scattering amplitude is not logarithmically enhanced. 
For increasing $y$, when BFKL evolution becomes important, the contribution of 
gluons with 
small virtualities to the amplitude decreases, and the suppression of double parton 
scattering is even stronger. 
Therefore the BFKL pomeron may be considered as a hard probe resolving the
proton structure at virtuality $|t|$.

The cancellation of the infrared divergencies is, in our approach,
represented by introducing a physical cutoff $s_0$ which is possible to
interpret in terms of impact factors. Here we shall rather view it as a
property of the confining QCD vacuum, in which propagation of gluons for
distances larger than about 1~GeV$^{-1}$ is suppressed \cite{GLUs0}. 
An approximate way to take this phenomenon into account is to modify 
the gluon propagator $1/k_i ^2 \to 1 /(k_i ^2 + s_0)$.

In the high energy limit, the differential cross-section for {\em quark-quark} scattering 
is given by 
\be
{d \hat\sigma_{qq} (\hat{s},t) \over  dt} = {|A(\hat{s},t)|^2 \over  16 \pi},
\label{sigqq}
\ee
where $\hat{s}$ is the squared invariant collision energy and $t$ the momentum transfer. 
For the case of scattering by colour singlet exchange at large 
momentum transfer $t$, but $\hat s \gg |t|$, the dominating, 
imaginary part of the amplitude is \cite{LIPATOV}
\begin{equation}
{\mathrm Im}\, A(\hat{s},t=-Q^2) = \int {d^2\tdm k \over (2\pi)^2} 
{\Phi^{ab}_0(\tdm k, \tdm Q) 
\Phi^{ab}(x,\tdm k,\tdm Q)\over
[(\tdm Q /2 + \tdm k) ^2 +s_0][(\tdm Q/2 - \tdm k)^2 +s_0]}.
\label{ima}
\end{equation}
Here, $x = |t|/\hat{s}$, and $\tdm k$ is defined so that the transverse momenta 
of the exchanged gluons are $\tdm k_{1,2} = \tdm Q /2 \pm \tdm k$, where 
$\tdm Q$ is the total transverse component of the momentum transfer $t=-Q^2$. 
The impact factor for {\em quark-quark} scattering
$\Phi^{ab}_0(\tdm k, \tdm Q)={(\delta^{ab}/ 2N_c)} \Phi_0(\tdm k, \tdm Q)$ 
describes the coupling of the quark to two gluons with adjoint colour indices $a,b$ 
in the colour singlet state, carrying the transverse momentum $\tdm Q$. 
The formulae for other relevant cases of {\em quark-gluon} and {\em gluon-gluon} scattering 
may be obtained from those given above by an appropriate modification of the colour 
charges.
Beyond the LL approximation when the running of the QCD coupling $g_s$ 
($\alpha_s = g^2_s / 4\pi$) has to be taken into account, one has 
$\Phi_0(\tdm k, \tdm Q)= g_s ((\tdm Q /2+\tdm k)^2 +s_0) g_s((\tdm Q /2-\tdm k)^2+s_0)$.
In the LL approximation, however, a fixed value of $g_s$ will be used.
$\Phi^{ab}$ is the result of the BFKL $x$-evolution from $\Phi_0 ^{ab}$
and is decomposed as
$\Phi^{ab} (x,\tdm k,\tdm Q) = {(\delta^{ab} /2N_c)}\Phi(x,\tdm k,\tdm Q)$.
The function $\Phi(x,\tdm k,\tdm Q)$ satisfies the following BFKL equation
%
\begin{eqnarray}   
\Phi(x,\tdm k,\tdm Q)& = & \Phi_0(\tdm k, \tdm Q)+ {3\alpha_s(\mu^2)\over
2\pi^2}\int_x^1{dx^{\prime}\over x^{\prime}} \int
{d^2\tdm k' \over (\tdm k' - \tdm k)^2 + s_0} \times
\nonumber \\ & & 
\left\{\left[{{\tdm k_1^2}\over {\tdm k_1^{\prime 2}} + s_0}   + 
{{\tdm k_2^2}\over {\tdm k_2^{\prime 2}} + s_0} 
  - Q^2 
 {(\tdm k' - \tdm k)^2+s_0 \over ({\tdm k_1^{\prime 2}} + s_0)
 ({\tdm k_2^{\prime 2}} + s_0)}  
\right] \times \right.
\nonumber \\ & & 
\Phi(x',\tdm k' ,\tdm Q) \, \Theta \left((k^2+Q^2/4)x'/x-k^{\prime 2} \right) 
\nonumber \\ & & 
- \left. \left[{{\tdm k_1^2}\over {\tdm k_1^{\prime 2}}  + 
(\tdm k' - \tdm k)^2 +2s_0} + 
{{\tdm k_2^2}\over {\tdm k_2^{\prime 2}}  + 
(\tdm k' - \tdm k)^2 +2s_0} \right] 
\Phi(x',\tdm k,\tdm Q) \right\}. 
\label{bfkl}
\end{eqnarray} 

This equation includes a treatment of the large next-to-leading corrections to
the BFKL equation. They have a complicated structure \cite{BFKLNL}, but their
dominant part may be approximately resummed to all orders by restricting the
integration region in the real emission term \cite{KC} which is equivalent to
resumming collinear divergencies in the NLL BFKL kernel \cite{COLLINEAR}.
Generalising to the case of a non-forward configuration with $Q^2 \ge 0$ the
relevant limitation is \cite{PSIPSI} 
\begin{equation} 
k'^2 \le (k^2+ Q^2/4) {x'\over x},
\label{kc}
\end{equation} 
which follows from  the requirement that the virtuality of the gluons exchanged
along the chain be dominated by the transverse momentum squared. The 
constraint (\ref{kc}) can be shown to exhaust about $70 \%$ of the
next-to-leading corrections to the QCD pomeron intercept \cite{BFKLNL,KC}. This
constraint is embodied in eq.~(\ref{bfkl}) by the step function 
$\Theta ((k^2+Q^2/4)x'/x-k^{\prime 2})$ which multiplies the real emission term.
Another part of the non-leading corrections may be accounted for by using the
running QCD coupling $\alpha_s(\mu^2)$ within the ladder, with the scale
$\mu^2$ related to the gluon virtualities. We use the one-loop QCD coupling
$\alpha_s$ 
with number of flavours $N_f=4$ and $\Lambda_{\rm\tiny QCD}=0.20 {\rm \; GeV}$,
with the scale $\mu^2=k^2+Q^2/4 + s_0$. Following our interpretation of the
parameter $s_0$, we have included it in eq.~({\ref{bfkl}}) in a way which
preserves gauge invariance. It is varied within the range $0.5 {\rm \; GeV}^2 <
s_0< 2 {\rm \; GeV}^2$. 
Eq.\ (\ref{bfkl}) reduces to the standard LL~BFKL equation after taking a fixed 
$\alpha_s$ (we set $\alpha_s=0.17)$ and substituting $\Theta(...) \to 1$. 
The equation is solved numerically in both cases; 
a summary of the numerical method and the adopted approximations is given in 
\cite{PSIPSI}. 

It is instructive to compare our approach with that of Mueller and Tang
\cite{MT}, who employ the exact solution of the non-forward LL~BFKL equation
\cite{LIPATOV}, given in terms of an infinite sum of contributions
labelled by the conformal spin~$n$. They retain only the $n=0$
eigenfunction of the BFKL kernel, which is the dominating contribution at
high rapidity $y$ and which has the anomalous dimension $\gamma = 1/2$. 
Therefore their amplitude for elastic parton-parton scattering is convergent in the
infrared and not sensitive to the details of the structure of the remnant.
The remnant is only necessary to ensure the colour neutrality of the 
scattering hadrons. At small rapidities $y$, however, the full set of 
BFKL eigenfunctions has to be taken into account because of the completeness 
relation and the non-suppression of components with higher conformal spins at 
$y=0$.

The contribution of non-zero $n$ at $y = 0$ may be estimated by 
comparing the amplitude for two-gluon exchange 
\begin{equation}
A_{2g}(t) = {8 \alpha_s^2 \over 9} \int d^2 \tdm k \; {1 \over 
(k_1 ^2 + s_0)(k_2 ^2 + s_0)}
= {16 \pi \alpha_s^2 \over 9 |t|} \log (|t|/s_0) +{\cal O}(s_0/|t|),
\label{2ge}
\end{equation}     
including the infrared cutoff, with the Mueller-Tang amplitude 
\begin{equation}
A_{MT}(y,t) = {32 \alpha_s ^2 \over 9 |t|} 
\int d\nu {\nu^2 \over (\nu^2 + 1/4)^2} e^{\omega(\nu)y}, 
\label{MT}
\end{equation}
where $\omega(\nu) = {3\alpha_s \over \pi} [2\psi(1) - \psi(1/2+i\nu) -\psi(1/2-i\nu)]$ and
$\psi(z)= \Gamma'(z) / \Gamma(z)$. The ratio 
\begin{equation}
{A_{2g}(t) \over A_{MT}(y=0,t) } \simeq {1\over 2} \log(|t|/s_0) 
\label{eq:a2g}
\end{equation}   
is a large number for sufficiently large $|t|$. This ratio illustrates a
mathematical property of the solutions to the BFKL equation, although at $y=0$
the high energy limit is strictly speaking a poor approximation. 

The large logarithm in (\ref{2ge}) is generated by asymmetric gluon 
configurations in which one of the exchanged gluons is soft with a 
small virtuality $k_1^2 \simeq s_0$, and the other is hard with
virtuality $k_2^2 \simeq Q^2$. 
The contribution of such two-gluon configurations to the colour 
singlet channel is suppressed for increasing $y$, 
because of an increasing probability of radiation from the hard gluon, 
which cannot be screened by the soft partner.
This phenomenon is embedded into the virtual corrections to the BFKL kernel. 
The virtual corrections to the gluon propagation may be 
exponentiated to give a {\em gluon reggeisation} 
suppression factor $(k_1^2 / Q^2)^{(3\alpha_s y / 2\pi)}$ \cite{Bartels,MMR}.
After including this factor into the integral in eq.~(\ref{2ge}) we obtain
an estimate of the non-asymptotic contribution of higher conformal
spins 
\begin{equation}
A_{na}(y,t)  
= {16 \pi \alpha_s^2 \over 9 |t|} 
{2\pi \over 3 \alpha_s y} \left[ 
1 - \left( {|t| \over e^2 s_0} \right)^{-(3\alpha_s y / 2\pi)} \right],
\label{eq:ahs}
\end{equation}     
where the $e^2$ factor was introduced in order to ensure
$A_{na}(y=0,t)+A_{MT}(y=0,t)=A_{2g}(t)$.

A recent analysis of the contributions from higher conformal spins \cite{MMR}
shows that, indeed, at smaller values of $y$ they become dominant and account
for the gluon reggeisation phenomenon. Thus, at moderate rapidities, such as
what is now experimentally 
accessible, the original Mueller-Tang result is not sufficient for a proper description
of the parton scattering amplitude at high $|t|$.
Let us mention, that the presence of the scale $s_0$ is reflected in 
the conformal representation by the appearance of an upper limit in the sum 
over conformal spins $n_{\rm\small max} \simeq (|t|/s_0)^{1/4}$ \cite{MMR}.

\begin{figure}
\begin{center}
\epsfig{file=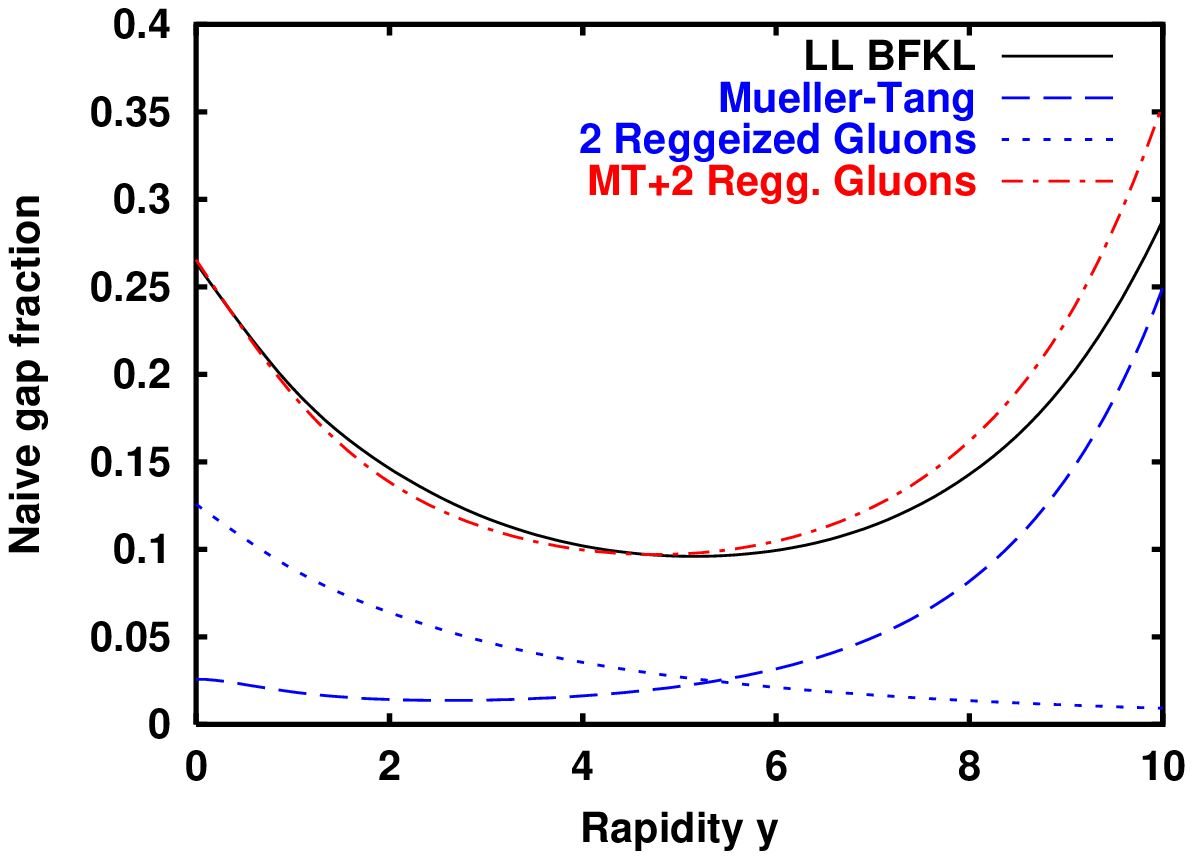, width=0.45\columnwidth}

\epsfig{file=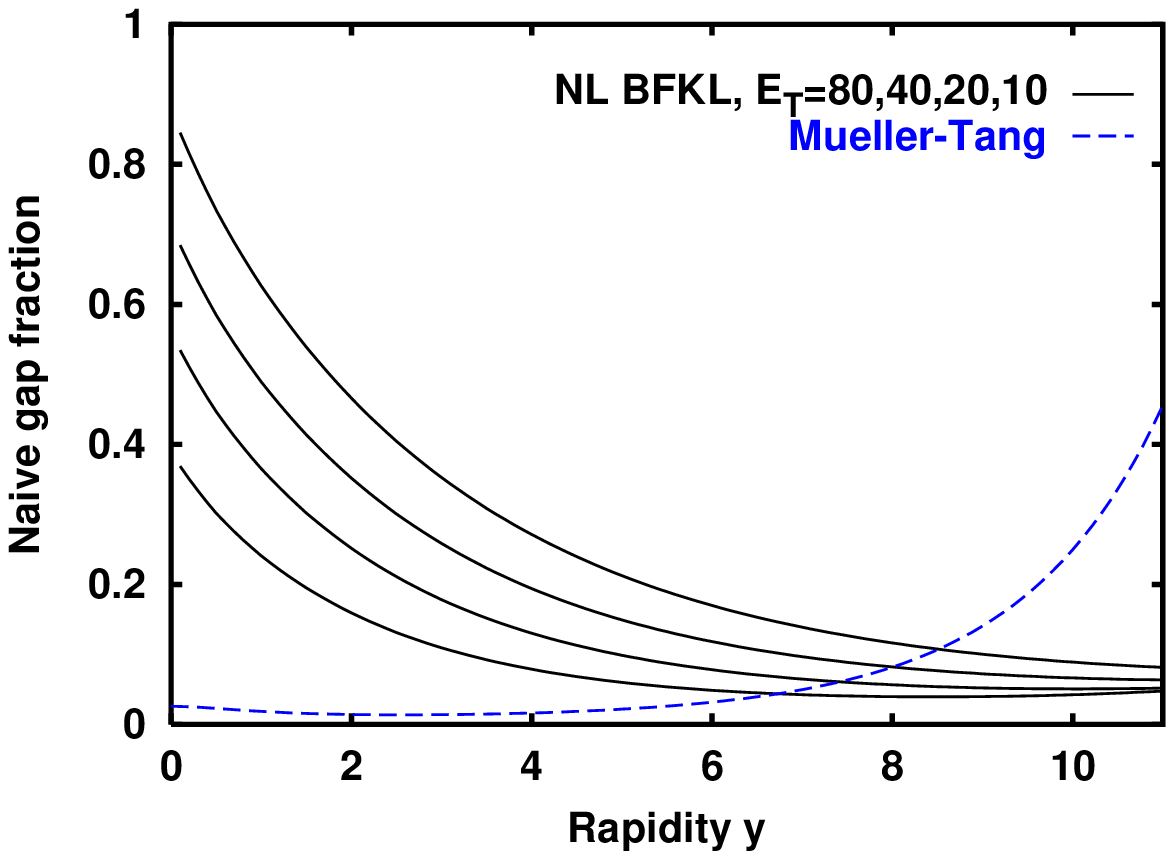, width=0.45\columnwidth}
\epsfig{file=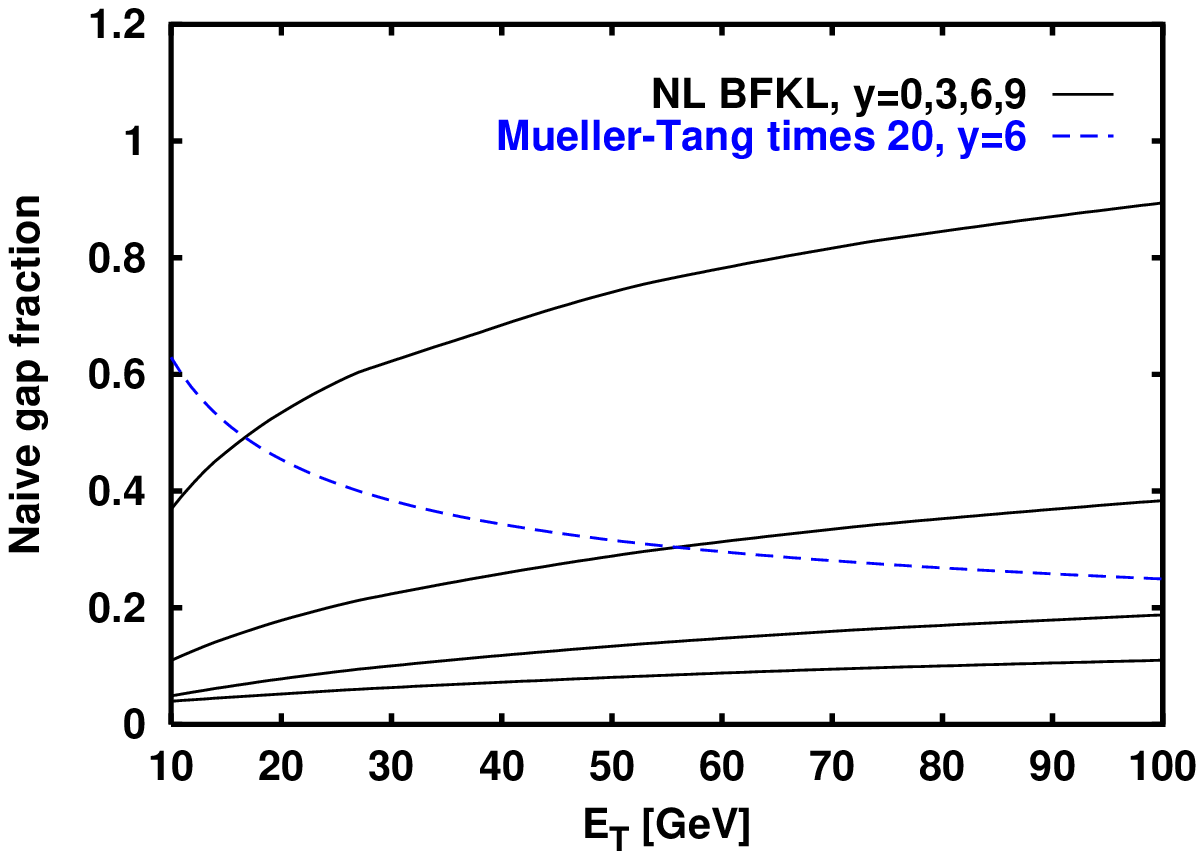, width=0.45\columnwidth}
\end{center}
\caption{Naive parton level gap fractions 
$[d\hat\sigma_{{\rm singlet}}(y,t)/dt] / 
[d\hat\sigma_{{\rm one-gluon}}(y,t)/dt]$ 
in quark-quark scattering via colour singlet exchange and via leading order
one-gluon exchange, versus the rapidity difference $y$ of the two scattered
partons and their transverse energy $E_T$.  
(a) Leading order BFKL solution (fixed $\alpha_s = 0.17$) for $E_T$=25~GeV compared with  
the Mueller-Tang asymptotic approximation to it, eq.\ (\ref{MT}), 
the reggeised gluon approximation eq.\ (\ref{eq:ahs}), 
and the sum of the two. (b,c) BFKL solution with non-leading corrections.}
\label{fig2}
\end{figure}

The ratio of the partonic differential cross-sections for colour singlet exchange and for 
leading order QCD quark-quark scattering by one-gluon exchange gives a naive estimate of 
the gap fractions before soft interactions and hadronisation are included.
In fig.\ \ref{fig2}a we compare the numerical results of the LL~BFKL equation, 
the Mueller-Tang equation (\ref{MT}), the non-asymptotic contribution corresponding to 
$A_{na}(y,t)$ as well as a simple sum of amplitudes $A_{na}(y,t)+A_{MT}(y,t)$. 
The last guess is surprisingly close to the full solution. 
The non-asymptotic contributions dominate up to $y=6$ despite the fact that the 
BFKL intercept  $1+\omega_0$ takes the rather high value of 1.45 which gives
a very rapid increase of the Mueller-Tang $n=0$ component.

Fig.\ \ref{fig2}bc shows the corresponding naive gap fractions 
for quark-quark scattering resulting from the BFKL equation with non-leading 
corrections (\ref{bfkl}) with $s_0=1\;{\mathrm GeV}^2$ and a running 
$\alpha_s(|t|)$ in $d\sigma_{{\rm one-gluon}}/dt$. 
The decrease of the gap fractions continuing up to values of $y$ 
larger than in the LL case is due to two effects. 
First, the non-leading corrections reduce the intercept and suppress the asymptotic, 
rising part. Second, the running coupling additionally enhances the contribution
of asymmetric configurations (with one of the gluons soft) and presumably
slows down the evolution of the reggeisation suppression factor with $y$.
Furthermore, the gap fraction rises with $E_T\simeq Q\simeq \sqrt{|t|}$ 
in spite of the fact that the running coupling was used. 
This is a specific property of the non-asymptotic part of the amplitude which 
still dominates even for $y \simeq 10$. One should note that these dependences on 
$y$ and $E_T$ of the gap fraction are in contrast to the predictions of the 
Mueller-Tang formula \cite{MT,D0DATA,CFL}. 
This shows the importance of both subleading $\log (1/x)$ corrections and 
the non-asymptotic effects in~$y$.
Finally, it was checked that a variation of $s_0$ between 0.5 and 2~GeV$^2$ does not
influence significantly the shapes in $E_T$ and $y$. However, it does
have a large effect on normalisation which increases (decreases) by a factor of 
about~1.5 when $s_0$ decreases from 1 to 0.5~GeV$^2$ (increases from 1 to 2~GeV$^2$)
in the relevant range of $y$ and $E_T$.

\section{Soft effects and Monte Carlo implementation}

In order to compare the perturbative results above with data, the parton level 
cross-section must be convoluted with the QCD evolved parton distributions 
$f_{i/p}(x,|t|)$ and $f_{j/\bar p}(x,|t|)$ of the incoming protons.
In the absence of soft rescattering the cross-section for production of two jets
initiated by partons $i$ and $j$ with a rapidity gap in between is given by
\begin{equation}
{d\sigma_{ij} \over dt\,dx_1\, dx_2} = 
\left( {C_i \over C_F} \right) ^2 x_1 f_{i/p} (x_1,|t|) \;
\left( {C_j \over C_F} \right) ^2 x_2 f_{j/\bar p} (x_2,|t|) \;
{d \hat\sigma_{qq} (x_1 x_2 s,|t|)\over dt}, 
\label{dsigij}
\end{equation}
where $s$ is the invariant \ppbar \ collision energy squared. 
The initial partons of species $i,j$ carry momentum fractions $x_{1,2}$ 
(cf.\ fig.~\ref{fig1}a) and have colour charges 
$C_{i,j}=C_F$ for quarks and $C_A$ for gluons. 

To consider rapidity gaps, which are defined as a region without final state
particles, one must also consider all additional activity in an event. This
means both pQCD processes, such as higher order parton emissions and multiple
parton scattering, as well as soft processes including hadronisation. A simple
way of doing this is to multiply the parton level cross-section
(\ref{dsigij}) with a gap survival probability factor $S^2$ and sum over all
parton species $i$ and $j$. This
probability may, however, depend on the subprocess with its associated colour
field topology, or on the kinematics in the event. For example, in the case of
gluon-gluon scattering or a larger momentum transfer, more partons are radiated
and a smaller gap survival probability is expected. 
The best way to handle these complex processes is via Monte Carlo event 
simulation. 

We have therefore implemented the obtained cross-section for elastic parton-parton scattering 
via colour singlet exchange, as well as the Mueller-Tang cross-section obtained from 
eq.\ (\ref{MT}), as new hard subprocesses in the event generator 
{\sc Pythia}.
Higher order parton emissions from the incoming and outgoing partons are
thereby included through conventional DGLAP parton showers and a model for
multiple interactions (MI) \cite{MI} in terms of additional perturbative
$2\to 2$ parton scatterings contribute to the underlying event complexity.
Such a description is, however, incomplete since it only accounts for
multiple exchanges of semihard gluons with $p_T^2 \gtrsim p^2_{T_0} \simeq
4\; {\mathrm GeV}^2$. 
Softer, non-perturbative interactions are modelled to take into account the
remnants of the interacting hadrons, which are extended objects, and the
hadronisation process through the Lund string model \cite{lund}. 

A successful approach to describe additional soft phenomena is given by the
soft colour interaction (SCI) model \cite{SCI}. 
Partons and remnants emerging from the hard process
can exchange colour-anticolour, representing soft gluons with negligible
momentum transfer. This leads to a modified colour string topology, producing 
a different final state after hadronisation.
For example, a rapidity region without strings may arise resulting in rapidity gap events, 
and the SCI model does indeed reproduce the diffractive hard scattering processed observed at
HERA \cite{SCIHERA} and the Tevatron \cite{SCI-TeV}. 
Moreover, a colour octet $c\bar{c}$ quark pair may be turned into a colour singlet reproducing 
charmonium production \cite{QQSCI,CRISTIANO}. 

The soft colour interaction model is included in the Monte Carlo event 
generator together with the multiple
interaction model in a way which is consistent with our previous study of 
diffractive hard scattering \cite{SCI-TeV}. 
As has been previously noted, SCI alone is not able to form 
sufficiently many gaps between jets when imposed on 
events with single hard gluon exchange \cite{SCI-jgj}.
The reason is that the soft interactions cannot screen 
emissions of gluons in the hard subprocess. 
Furthermore, contrary to the perturbative semihard gluons in eq.\ (\ref{2ge}), 
the SCI contribution is not enhanced by a loop integration. 
In the present case of gaps between jets, the SCI has 
the effect of rearranging strings to span across 
the parton level gap from the BFKL singlet exchange and therefore 
it reduces the gap survival probability. 
 
\section{Comparison with data}
Gaps between jets have been observed in \ppbar \ collisions at
$\sqrt{s}=1800$~GeV by the D\O{} \cite{D0DATA} and CDF \cite{CDFDATA}
collaborations. The results are presented in terms of the fraction of two-jet
events that have a rapidity region $|\eta | < 1$ with no particles in
between the jets. The jets are reconstructed using a cone algorithm and have
transverse energy $E_T$ defining different samples. D\O{} has a low-$E_T$
sample with $15 < E_{T2} < 25$ GeV, and a high-$E_T$ sample with $E_{T2} > 30$
GeV, where $E_{T2}$ is the second highest jet transverse energy. CDF has one
sample with $E_{T} > 20$ for both jets. 

The same criteria and analyses as for the data can be applied to the Monte Carlo
events generated 
with the parton-level BFKL cross-section (including non-leading corrections). 
The gap fraction is given by the ratio of hard colour singlet exchange events 
to the standard LO QCD processes. 
The effects of the underlying event leading to the destruction of rapidity gaps
are studied in detail and illustrated by three different treatments. The first
and simplest method is to multiply the generated cross-section for gaps at the
parton level with a constant survival probability $S^2$ taken as a free parameter.
The second method is to use the MI model 
(in its most sophisticated version given by setting {\tt MSTP(82)=4} in  \pythia ) 
followed by standard hadronisation. 
The third and most elaborate method is to use both MI and SCI before applying string hadronisation. 
The value of the infrared regulator of the MI model $p_{T_0}$ has here been increased to avoid 
double counting of soft interactions, as described in \cite{SCI-TeV}.

\begin{figure}
\begin{center}
\epsfig{file=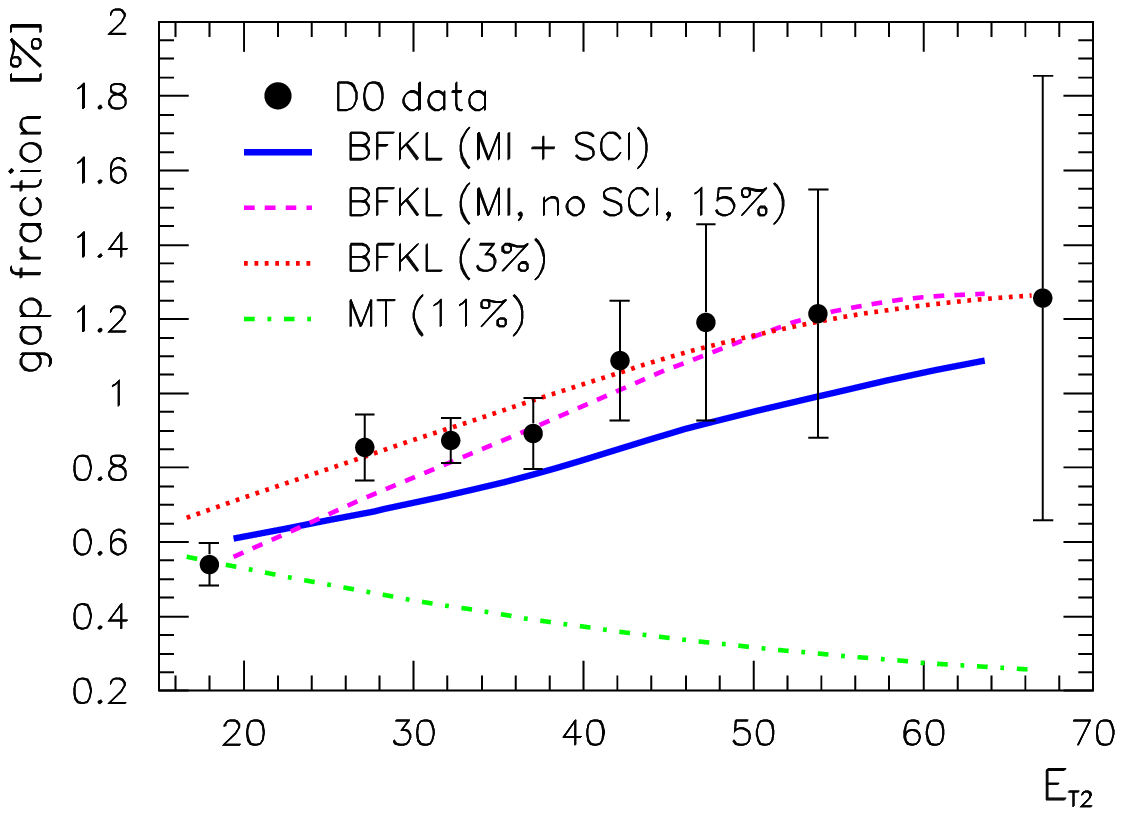, width=0.45\columnwidth}

\epsfig{file=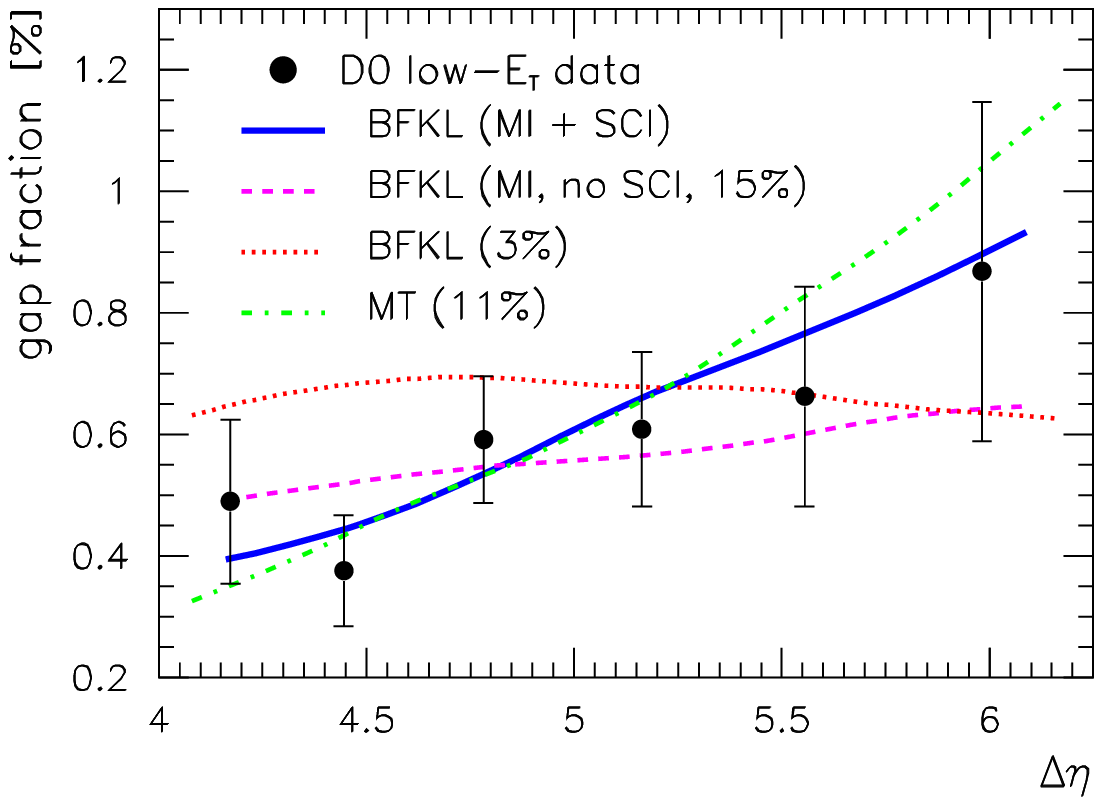, width=0.45\columnwidth}
\epsfig{file=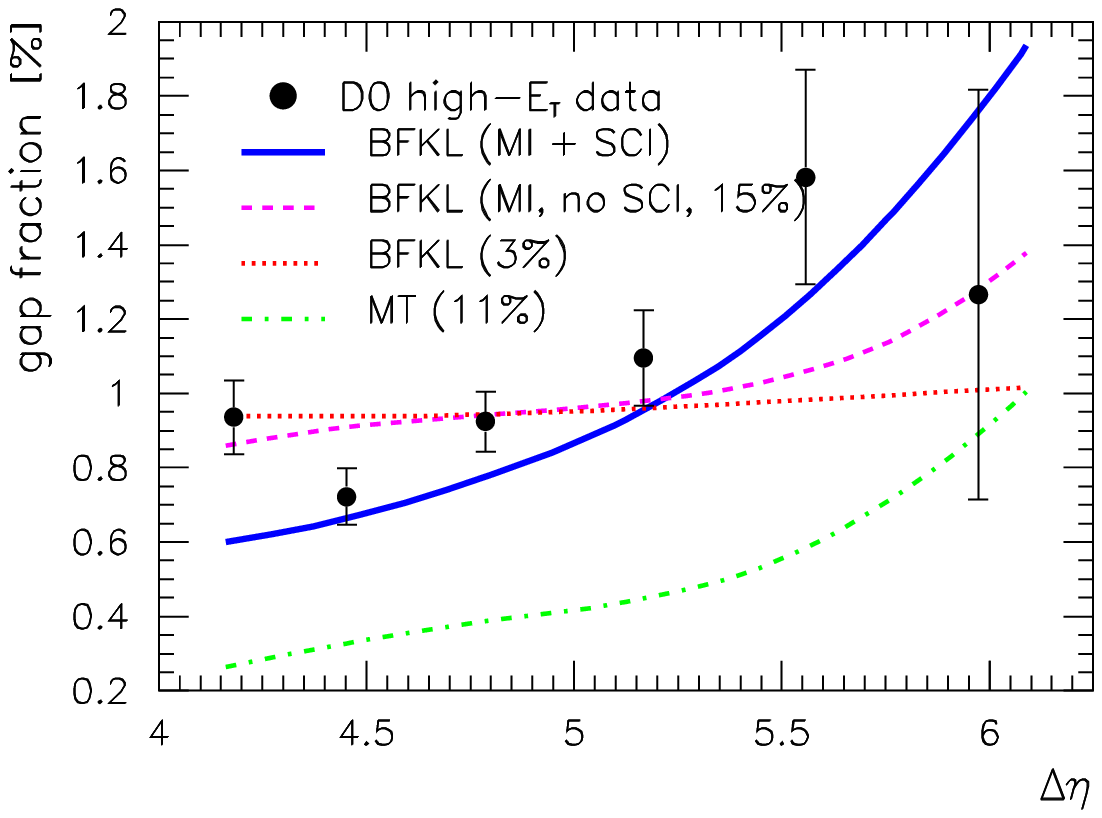, width=0.45\columnwidth}
\end{center}
\caption{
Fraction of jet events having a rapidity gap in $|\eta| < 1$ between the jets,
versus (a) the second highest jet-$E_T$ and (b,c) the rapidity separation 
$\Delta\eta$ between the jets for low and high jet-$E_T$ ($15<E_{T2}<25$~GeV
and $E_{T2}>30$~GeV). D\O{} data \protect\cite{D0DATA} compared to the colour 
singlet exchange mechanism based on BFKL equation 
with non-leading corrections with the underlying event treated in three ways:
simple 3\% gap survival probability, multiple interactions (MI) and
hadronisation requiring a 15\% gap survival probability, MI plus soft colour
interactions (SCI) and hadronisation with no need for an overall
renormalisation factor. Also shown is the Mueller-Tang (MT) calculation with an
11\% gap survival probability.}
\label{fig3}
\end{figure}

In fig.\ \ref{fig3} we compare the model results with the D\O{} data. 
First, one can notice that the Mueller-Tang approximation with constant gap survival
 probability gives a wrong $E_T$ dependence of the gap fraction as well as a somewhat 
too large slope of the rapidity dependence. Going beyond
the large-$y$ approximation and including the 
non-leading BFKL corrections, gives an improved result that can reproduce 
the data quite well. Without our most elaborate model of the underlying event, 
however, one needs an overall renormalisation in terms of a gap survival 
probability. Since no such factor is needed when both the multiple 
interaction model and the soft colour interaction model are included, 
it seems that these models account properly for the gap survival 
probability. The normalisation is here rather sensitive to the value of the 
probability parameter $P$ for a soft colour exchange between partons. For 
example, the gap cross-section decreases by a factor two when increasing $P$ 
from 0.20 to 0.30. In contrast, the shapes of the $y$ and $E_T$-distributions 
are not sensitive to variations of $P$.
We have used the value $P=0.25$, which is within the range 
allowed by the previous analyses of the SCI model discussed above.

One should note that the solution to the BFKL equation 
with non-leading corrections gives a shape of the $E_T$ distribution (fig.~\ref{fig3}a), 
which is stable against variations of the model of the  rescattering process in
the underlying event. This gives strong support for the BFKL approach and shows
that the Mueller-Tang approximation is not valid in the kinematical range of
the Tevatron. The $\Delta\eta$ dependence is much more sensitive to
the treatment of rescattering processes. The inclusion of multiple interactions
slightly increases the slope of the rapidity dependence and an even stronger
increase is observed when the SCI model is also turned on. 

\begin{figure}
\begin{center}
\epsfig{file=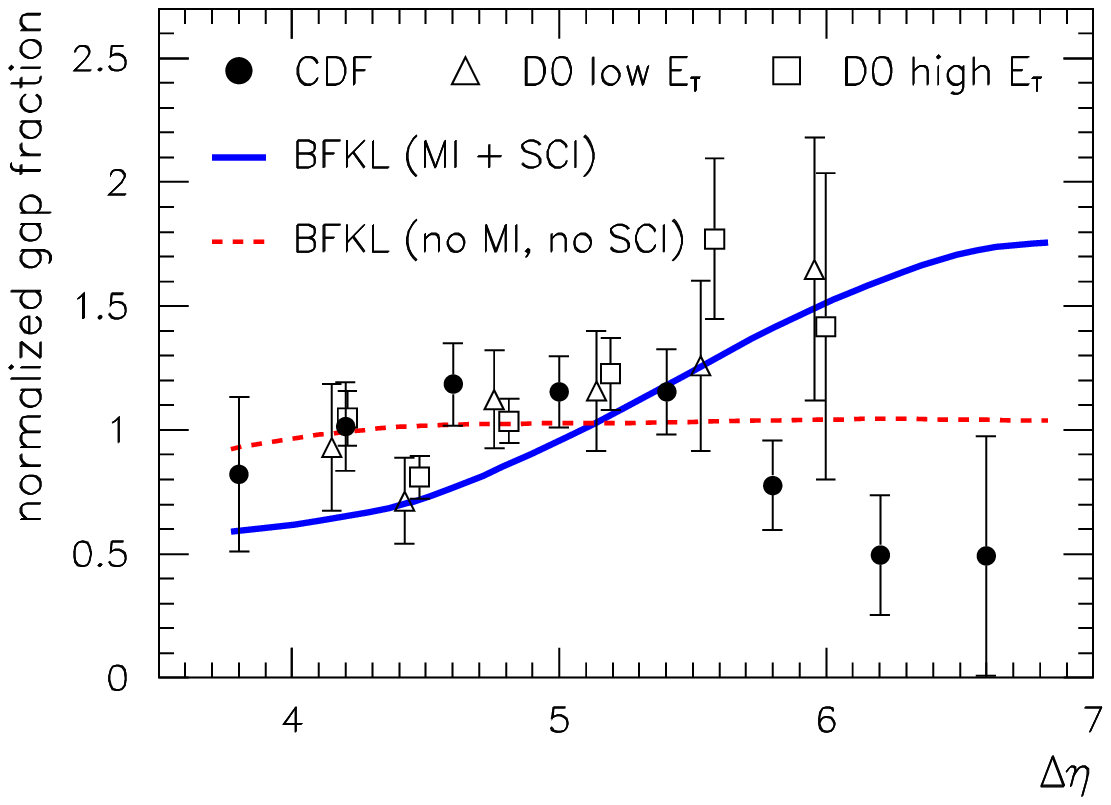, width=0.45\columnwidth}
\epsfig{file=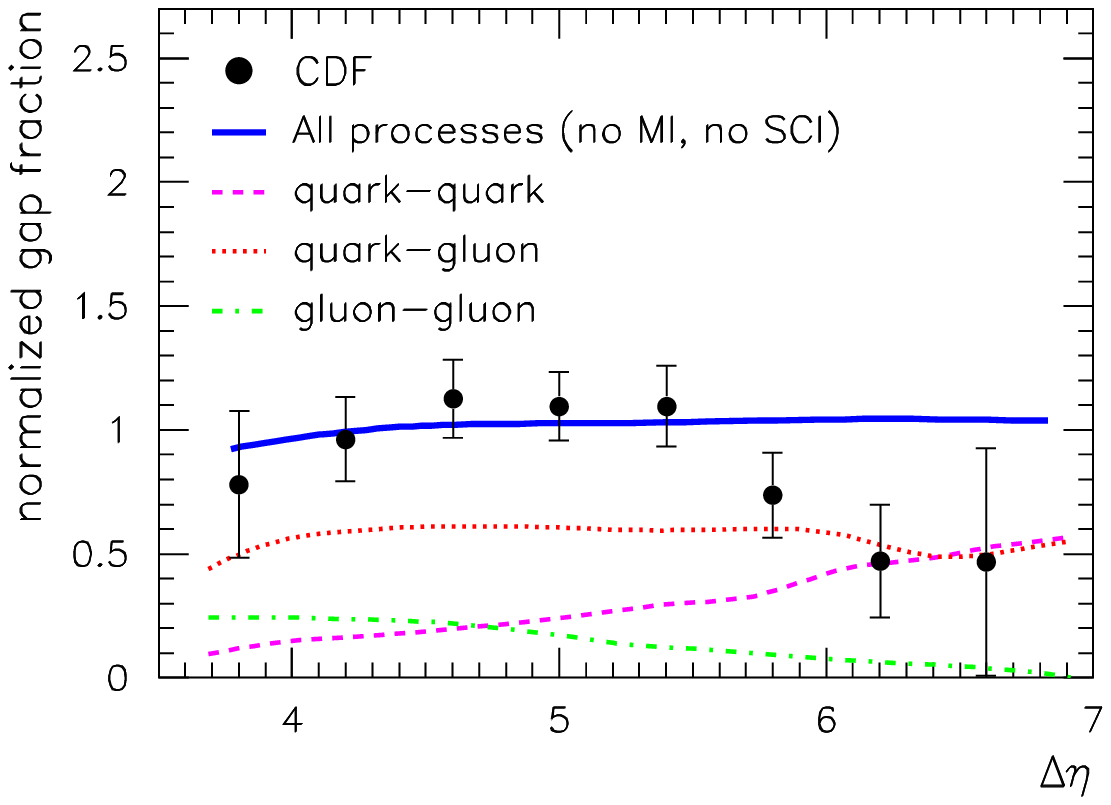, width=0.45\columnwidth}
\end{center}
\caption{Gap fraction as in fig.~\ref{fig3}, but normalised to the average gap
fraction, versus the rapidity separation $\Delta\eta$ between the jets. 
(a) The BFKL colour singlet exchange model with (solid curve) and without
(dashed curve) underlying event simulation, calculated with conditions as for
the shown CDF data \protect\cite{CDFDATA}. 
Rescaled D\O{} data from fig.~\ref{fig3} are included for comparison. 
(b) BFKL result for the contributing subprocesses of $qq$, $qg$ and $gg$
scattering, but without multiple interactions and soft colour interactions.}
\label{fig4}
\end{figure}

The CDF data for the gap fraction as a function of $\Delta\eta$ are shown in
fig.\ \ref{fig4}. A discrepancy with the BFKL+MI+SCI model can be observed
in the region $\Delta\eta > 5.5$. The CDF data show a different tendency
than the D\O{} data, but because of the large errors it cannot be claimed 
that the data are incompatible.
The BFKL result with constant gap survival probability 
$S^2 = 3\%$ falls between 
these data sets and therefore gives a reasonable overall description.
Recall however, that in this case $S^2$ is arbitrary.

In fig.\ \ref{fig4}b, this model is decomposed into the contributions from the
different hard scattering subprocesses. Quark-gluon scattering dominates except
at the largest $\Delta\eta$ where quark-quark scattering contributes equally,
whereas gluon-gluon scattering is only important at the lowest $\Delta\eta$. 
These relative contributions can be understood, since both gluon and quark
initiated processes contribute to the cross-section (\ref{dsigij}) in these jet
samples where the initial partons typically have a fraction $x \sim 0.1$ of the
beam momentum. The gluon contribution 
in the HCS exchange is enhanced in relation to the quark one by $(C_A/C_F)^2$,
the ratio of the colour charges squared. 
This partially compensates for the smaller value of the gluon distribution in 
the probed region of $x$. 
For $\Delta \eta \sim 4$, $x$ may be smaller and one has more gluon scattering, 
whereas for $\Delta \eta \sim 6$ quark scattering dominates.  

The increase of the rapidity slope of the gap fraction caused by the SCI model
can now be traced. The colour string topology in the Lund model depends on which
parton that was scattered. A scattered valence quark leaves a diquark remnant
as endpoint of a string, whereas a scattered gluon leaves a diquark plus a
quark stretching two strings to the remnant. Thus, with a constant probability
for colour exchange within parton pairs in the SCI model, there is a larger 
probability for string
reconnections with the more partons and strings available in case of gluon
scattering. For the hard colour singlet exchange this means a larger
probability to get a string across the parton level gap in case of gluon
scattering. The gap survival probability is therefore smaller for gluon-gluon
scattering, being more frequent at small $\Delta\eta$, and largest for
quark-quark scattering at larger $\Delta\eta$. 

Finally, let us recall that the gap fraction is defined relative to the 
total rate of high-$E_T$ jet pairs with the same rapidity separation. 
This corresponds to production of Mueller-Navelet jets \cite{MN}, 
which is not well understood when the same $E_T$-cut is imposed on both jets. 
The experimental results suggest a rather steep dependence of the
Mueller-Navelet cross-section on the rapidity separation between the jets \cite{MNEXP}. 
Recent theoretical considerations \cite{MNTHEO} indicate a 
BFKL-driven increase of the cross-section for production of Mueller-Navelet jets 
(with $E_{T\; cut}=20$~GeV) by about 50\% when $\Delta \eta$ increases from 4 to~6.
However, the equal cut on $E_T$ makes this observable infrared unsafe, \ie,
unstable against variations of the treatment of higher order QCD corrections. 
It is still fair to say that both theory and experiment give an expectation of 
an increase with rapidity of the denominator of the gap fraction.
This would, of course, have some impact on the $\Delta \eta$ shape of our 
predictions. We have used {\pythia} with conventional $2\to 2$ QCD matrix 
elements for the denominator in the gap fraction. Inclusion of the suggested 
increasing tendency would affect our model results and reduce their 
$\Delta y$-slope. This would bring the results obtained with MI+SCI in 
better agreement with the CDF data in fig.~\ref{fig4}a, whereas the simpler 
model with constant gap survival probability would get a negative slope that 
spoils its agreement with data. Lacking a reliable theoretical understanding 
of the Mueller-Navelet jet production mechanism, we leave this problem open for
future studies. 

\section{Conclusions}

We have investigated the production of high-$E_T$ jets separated by a central
rapidity gap as observed in $p\bar p$ collisions at the Tevatron. The data can
be understood in terms of a model based on
hard parton scattering through the exchange of a colour singlet gluon ladder described by the
BFKL equation. The two-gluon exchange amplitude with resummed virtual corrections 
plays a crucial role in the Tevatron kinematic range, where the energy is not asymptotically 
large. This makes the previously used Mueller-Tang approximation inadequate. 
The account for non-asymptotic effects in $y$ and non-leading corrections in 
the BFKL formalism improves the theoretical understanding and results in
distributions in $E_T$ and $\Delta\eta$ of the parton level gap fraction in
much better qualitative agreement with the data. 
Our implementation of the partonic BFKL process in the {\pythia} Monte Carlo
provides a detailed treatment of soft dynamics giving a gap survival
probability that varies event by event. 
The full event simulation is particularly important for the rapidity gap
observable defined by the final state properties. 

A detailed comparison with the D\O{} and CDF data shows an overall good agreement. 
In particular, the model is able to reproduce the magnitude of the gap fraction. 
The $E_T$ distribution of the gap fraction coincides with the experimental results. 
This conclusion holds both at the partonic and hadronic level 
regardless of the treatment of soft effects.
On the other hand, our result for the rapidity dependence of the gap fraction 
is somewhat uncertain. This problem is not related to the hard colour singlet 
exchange  but to an unresolved issue of rapidity dependence of the cross-section for 
the production of Mueller-Navelet jets. 
Within this uncertainty, a reasonable description is obtained for the 
$\Delta \eta$ distribution of the gap fraction.

In conclusion, the BFKL approach to hard colour singlet scattering gives a 
good understanding of the jet-gap-jet events observed at the Tevatron. 
This demonstrates the importance of BFKL dynamics in QCD. \\

\noindent
{\bf Acknowledgements:}
We are grateful to J.~Kwieci\'{n}ski and M.~Ryskin for helpful discussions
and to J.~Rathsman and N.~T\^{\i}mneanu for critical reading of the manuscript.
This research was partially supported by the Swedish Natural Science Research Council
and by the Polish Committee for Scientific Research (KBN) grant no.\ 5P03B~14420.



\pagebreak

\end{document}